\documentclass[article,reprint,amsmath,amssymb,superscriptaddress,longbibliography]{revtex4-1}

\usepackage{graphicx}
\usepackage{dcolumn}
\usepackage{bm}
\usepackage{hyperref} 
\usepackage{breakurl}
\usepackage{multirow}
\usepackage{enumitem}
\usepackage{color}
\usepackage[pagewise]{lineno}
\setcitestyle{super} 

\begin{document}

\title {Anisotropic superconductivity in topological crystalline metal Pb$_{1/3}$TaS$_2$ with multiple Dirac fermions}

\author{Xiaohui Yang}
   \thanks{Equal contributions}
   \email{yangxiaohui@westlake.edu.cn}
     \affiliation{Institute of Natural Sciences, Westlake Institute for Advanced Study, Hangzhou 310024, P. R. China}
      \affiliation{Key Laboratory for Quantum Materials of Zhejiang Province, School of Science, Westlake University, Hangzhou 310024, P. R. China}

\author{Tonghua Yu}
\thanks{Equal contributions}
      \affiliation{Department of Applied Physics, University of Tokyo, Tokyo 113-8656, Japan}
\author{Chenchao Xu}
      \affiliation{Department of Applied Physics, University of Tokyo, Tokyo 113-8656, Japan}
\author{Jialu Wang}
      \affiliation{Institute of Natural Sciences, Westlake Institute for Advanced Study, Hangzhou 310024, P. R. China}
      \affiliation{Key Laboratory for Quantum Materials of Zhejiang Province, School of Science, Westlake University, Hangzhou 310024, P. R. China}
\author{Wanghua Hu}
      \affiliation{Institute of Natural Sciences, Westlake Institute for Advanced Study, Hangzhou 310024, P. R. China}
      \affiliation{Key Laboratory for Quantum Materials of Zhejiang Province, School of Science, Westlake University, Hangzhou 310024, P. R. China}

\author{Zhuokai Xu}
      \affiliation{Institute of Natural Sciences, Westlake Institute for Advanced Study, Hangzhou 310024, P. R. China}
      \affiliation{Key Laboratory for Quantum Materials of Zhejiang Province, School of Science, Westlake University, Hangzhou 310024, P. R. China}
 \author{Tao Wang}
      \affiliation{Institute of Natural Sciences, Westlake Institute for Advanced Study, Hangzhou 310024, P. R. China}
      \affiliation{Key Laboratory for Quantum Materials of Zhejiang Province, School of Science, Westlake University, Hangzhou 310024, P. R. China}
 \author{Chao Zhang}
      \affiliation{Instrumentation and Service Center for Physical Sciences, Westlake University, Hangzhou 310024, China}
\author{Zhi Ren}
      \affiliation
      {Institute of Natural Sciences, Westlake Institute for Advanced Study, Hangzhou 310024, P. R. China}
      \affiliation{Key Laboratory for Quantum Materials of Zhejiang Province, School of Science, Westlake University, Hangzhou 310024, P. R. China}

\author{Zhu-an Xu}
      \affiliation{Zhejiang Province Key Laboratory of Quantum Technology and Device, Department of Physics, Zhejiang University, Hangzhou 310027, P. R. China}
      \affiliation{State Key Lab of Silicon Materials, Zhejiang University, Hangzhou 310027, P. R. China}

\author{Motoaki Hirayama}
      \affiliation{Department of Applied Physics, University of Tokyo, Tokyo 113-8656, Japan}
     \affiliation{RIKEN Center for Emergent Matter Science, 2-1 Hirosawa, Wako, 351-0198, Japan}

\author{Ryotaro Arita}
     \affiliation{Department of Applied Physics, University of Tokyo, Tokyo 113-8656, Japan}
     \affiliation{RIKEN Center for Emergent Matter Science, 2-1 Hirosawa, Wako, 351-0198, Japan}

\author{Xiao Lin}
\email{linxiao@westlake.edu.cn}
     \affiliation{Institute of Natural Sciences, Westlake Institute for Advanced Study, Hangzhou 310024, P. R. China}
      \affiliation{Key Laboratory for Quantum Materials of Zhejiang Province, School of Science, Westlake University, Hangzhou 310024, P. R. China}

\date{\today}

\begin{abstract}

Topological crystalline metals/semimetals (TCMs) have stimulated a great research interest, which broaden the classification of topological phases and provide a valuable platform to explore topological superconductivity.
Here, we report the discovery of superconductivity and topological features in Pb-intercalated transition-metal dichalcogenide Pb$_{1/3}$TaS$_2$.
Systematic measurements indicate that Pb$_{1/3}$TaS$_2$ is a quasi-two-dimensional (q-2D) type-II superconductor ({\em T}$_c \approx$ 2.8 K) with a significantly enhanced anisotropy of upper critical field ($\gamma_{H_{c2}}$ = $H_{c2}^{ab}/H_{c2}^{c}$ $\approx$ 17).
In addition, first-principles calculations reveal that Pb$_{1/3}$TaS$_2$ hosts multiple topological Dirac fermions in the electronic band structure. We discover four groups of Dirac nodal lines on the $k_z = \pi$ plane and two sets of Dirac points on the rotation/screw axes, which are protected by crystalline symmetries and robust against spin-orbit coupling (SOC). Dirac-cone-like surface states emerge on the (001) surface because of band inversion.
Our work shows that the TCM candidate Pb$_{1/3}$TaS$_2$ is a promising arena to study the interplay between superconductivity and topological Dirac fermions.

\end{abstract}

\maketitle

\noindent\textbf{Introduction}

\noindent The search for exotic topological phases of condensed matter has attracted a significant attention since the discovery of topological insulators (TIs), a unique class of electronic systems that show insulating bulks and topologically protected boundary excitations \cite{Bi2Te3,moore2010birth,TI2010,ando2013topological,Na3Bi,TaAs2015,YbMnBi2,PbTaSe22016topological,ZrSiS2016}. Shortly following TIs, topological metals/semimetals (TSMs) with bulk band crossings close to the Fermi level are broadly proposed and verified \cite{PhysRevB.83.205101,PhysRevB.85.195320,TSMs,DSMs,burkov2016topological,hirayama2017topological,hirayama2018topological,fujioka2019strong,yamada2019large}. Because of the topological bulk and surface states, for instance, nodal-line metals/semimetals (NLSMs) exhibit unconventional transport features, such as three-dimensional quantum Hall effect (3D QHE) and high-temperature surface superconductivity \cite{3DQHE,HighTc}. In the absence of magnetism, a centrosymmetric TSM, where both spatial inversion ($\mathcal{P}$) and time reversal ($\mathcal{T}$) symmetries are preserved, may host quadruply degenerate bulk nodes resembling massless Dirac fermions \cite{DSMs,PhysRevB.85.195320}. Compared with the noncentrosymmetric case, however, these degeneracies are unstable under significant spin-orbit coupling (SOC) unless extra crystalline symmetries are present \cite{PhysRevB.83.205101}. To be more specific, apart from the $\mathcal{PT}$ protection, a stable Dirac nodal point entails the guarantee of a rotation or screw symmetry \cite{DSMs,PhysRevB.85.195320}, and a Dirac nodal line can be supported by additional nonsymmorphic operations \cite{PhysRevLett.115.126803,PhysRevB.92.081201,PhysRevB.95.075135,PhysRevB.97.045131}. Crystalline symmetries are therefore an essential factor for the search of TSMs hosting stable bulk Dirac fermions. We alternatively refer to such TSMs as topological crystalline metals/semimetals (TCMs) \cite{chen2015topological}.

Recently, the family of the so-called $112$ systems, $MTX_2$ ($M$ = Pb, Sn, Tl or In, $T$ = Ta or Nb, $X$ = Se or S), have created a surge of research activities, because of the superconductivity and rich topological nature \cite{PbNbSe22016,TlTaSe2,InxTaSe2,InxTaS2,InNbS2}. The noncentrosymmetric PbTaSe$_2$ (derived from $1H$-TaSe$_2$ by intercalating Pb in the van der Waals gap) was reported to be a promising topological superconductor (TSC) candidate, due to the observation of zero-energy Majorana bound states in the vortices \cite{Pb112TSS}. PbTaSe$_2$ is also a typical NLSM where the nodal lines are guaranteed by mirror symmetry \cite{PbTaSe22016topological}. A different group of $112$ systems with centrosymmetric lattices, e.g., PbTaS$_2$ \cite{PbTaS2} and SnTaS$_2$ \cite{SnTaS2PRB}, in which the nodal lines are protected by the $\mathcal{PT}$ symmetry instead, without the account of SOC. Nevertheless, nodal lines in the centrosymmetric $112$ systems receive no nonsymmorphic protection and thus cannot survive under the strong SOC \cite{SnTaS2PRB}.

Herein, based on symmetry analysis and band calculations, we predict a new TCM candidate, the centrosymmetric crystal Pb$_{1/3}$TaS$_2$ that hosts multiple stable nodal point and line structures. In contrast to the above-mentioned $112$ systems, nodal lines in Pb$_{1/3}$TaS$_2$ are robust against SOC by virtue of the nonsymmorphic symmetry. Dirac points are furthermore unveiled, stabilized by rotation/screw symmetries. Band inversion gives rise to Dirac-cone-like surface states in the (001) surface. In addition, we synthesize and characterize this single crystal in experiment. The results show that Pb$_{1/3}$TaS$_2$ peculiarly exhibits quasi-two-dimensional (q-2D) superconductivity ($T_c$ = 2.8 K) with large anisotropy in upper critical field $H_{c2}$. Consequently, the superconductor Pb$_{1/3}$TaS$_2$ is a unique electronic system manifesting versatile nontrivial topological nature, offering a realistic material testbed for the exploration of the Dirac fermions and even possible TSC.

\vspace{3ex}
\noindent\textbf{Results}

\begin{figure}[bhtp]  
\begin{center}
\includegraphics[width=3.2in]{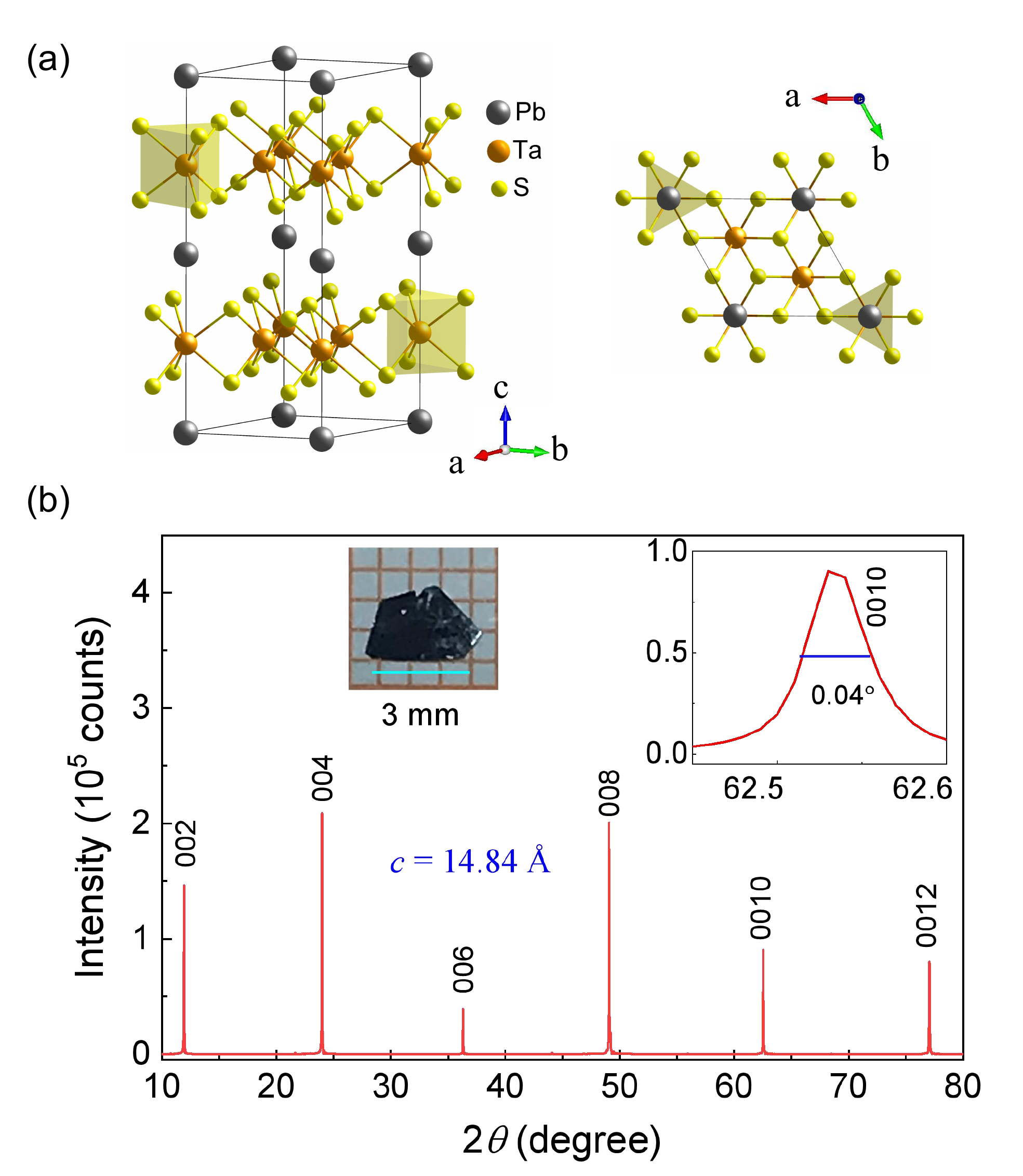}
\end{center}
\caption{\label{Fig1}(a) The crystal structure of Pb$_{1/3}$TaS$_2$ along different directions. (b) XRD pattern of the single crystal with (00$l$) reflections, the inset of the right panel zooms in the (0010) reflection.}
\end{figure}

\noindent\textbf{Sample characterizations}
Pb$_{1/3}$TaS$_2$ is centrosymmetric with a hexagonal structure and a space group $P6_3/mcm$ (No. 193) \cite{fang1996crystal}. As illustrated in Fig. \ref{Fig1}(a), Ta atoms are in trigonal-prismatic coordination by S atoms and the stacking sequence of S-Ta-S sandwiches follows that in 2$H$-TaS$_2$\cite{polytype2004}. Pb atoms are intercalated in between TaS$_2$ layers, but the occupation number is only one third of that in $112$ PbTaS$_2$ phase\cite{PbTaS2}.
Fig. \ref{Fig1}(b) presents the XRD pattern of Pb$_{1/3}$TaS$_2$ single crystals normal to ab-plane. The inset shows the full width at half-maximum (FWHM) of the (0010) peak is only 0.04$^{\circ}$, indicating the high crystalline quality. The interplanar spacing is calculated to be 14.84 \r{A} by employing the least-square method.
According to the EDX data (see Fig. S1 of Supporting Information), the molar-ratio between Pb, Ta and S atoms amounts to 1: 3: 6, in good agreement with the nominal one.

\begin{figure*}[htpb]
\centering
\includegraphics[width=6.2in]{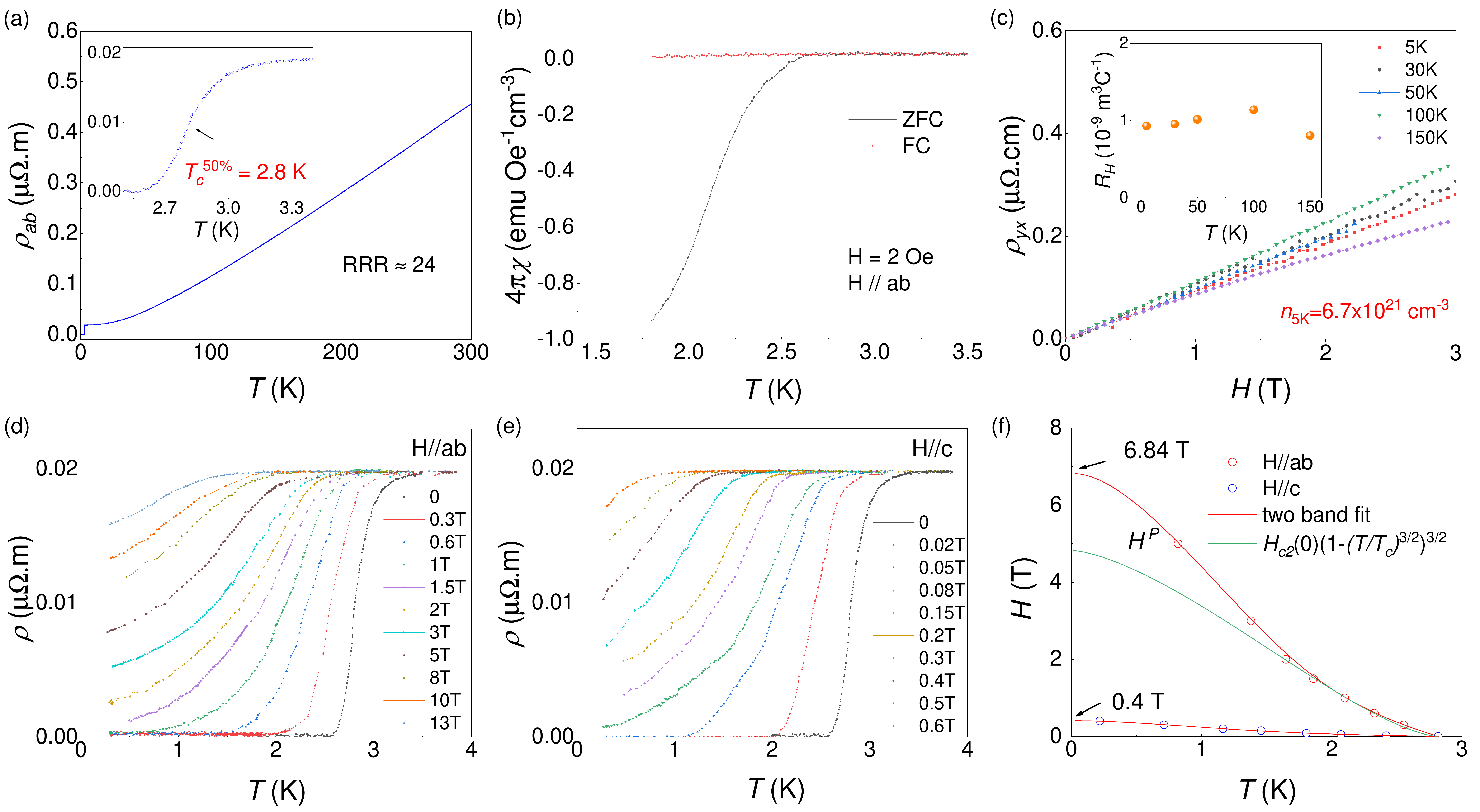}
\caption{\label{Fig2}(a) $T$ dependence of the electrical resistivity
${\rho}_{ab}$ of Pb$_{1/3}$TaS$_2$ single crystal. The inset shows the superconducting transition around $T_c$. (b) $T$ dependence of dc magnetic susceptibility ($H$//$ab$, $H$ = 2 Oe) around $T_c$. (c) Magnetic field dependence of Hall resistivity for Pb$_{1/3}$TaS$_2$ at different temperatures. Inset: the Hall coefficient vs. temperature.
(d) and (e) The low $T$ resistivity under different magnetic fields of single crystal, magnetic field parallel and perpendicular to the ab-plane, respectively. (f) $T$ dependence of the $H_{c2}$ with two-band fits for both directions, the black line dictates the Pauli paramagnetic limit.}
\end{figure*}

\begin{figure}[htpb]
\centering
\includegraphics[width=3.5in]{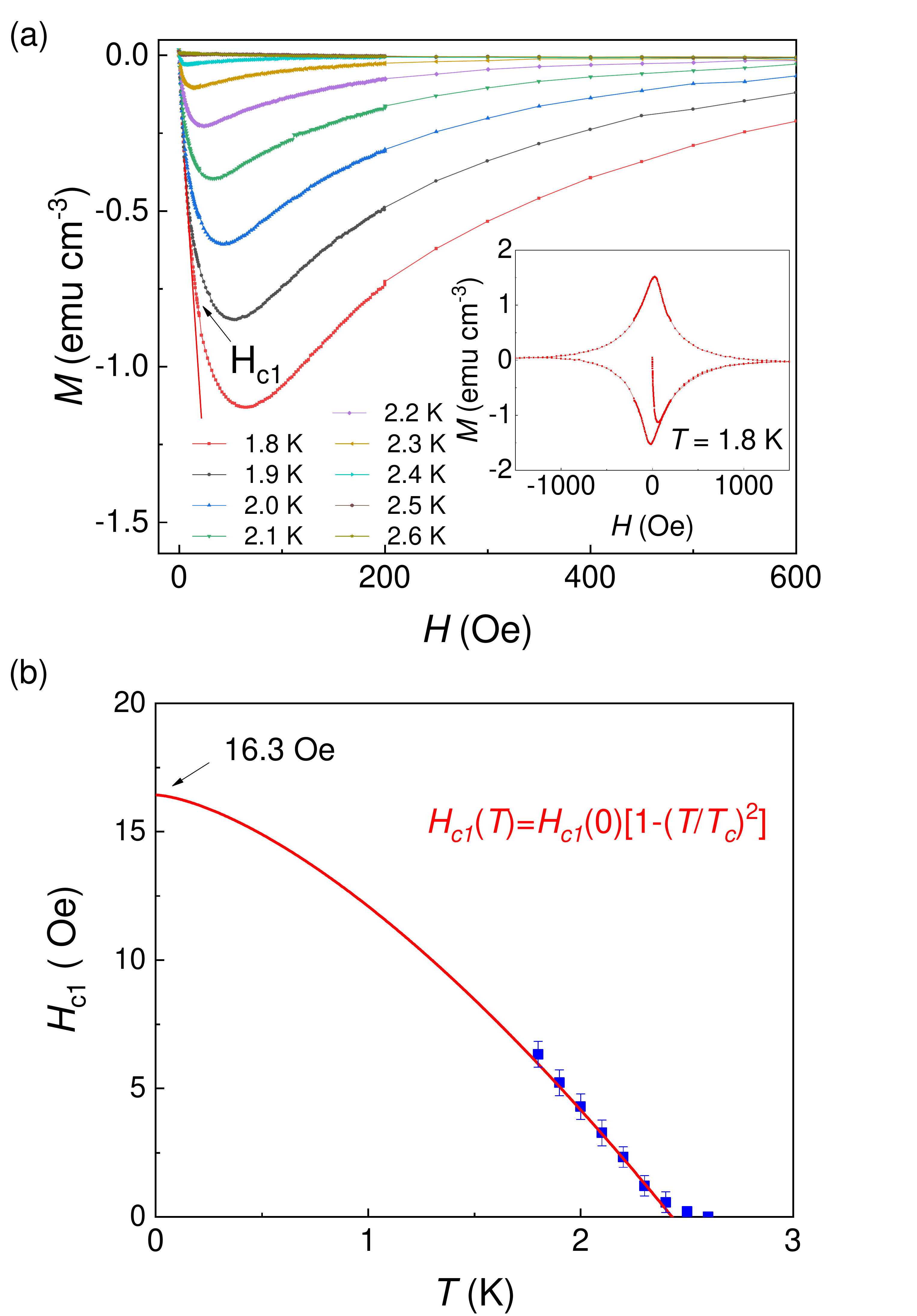}
\caption{\label{Fig3}(a) Magnetization $M(H)$ curves at various temperatures of Pb$_{1/3}$TaS$_2$ single crystal, the inset shows the loop taken at 1.8 K. (b) The superconducting $H$-$T$ phase diagram of Pb$_{1/3}$TaS$_2$.}
\end{figure}

\begin{figure*}
    \centering
    \includegraphics[width=6.2in]{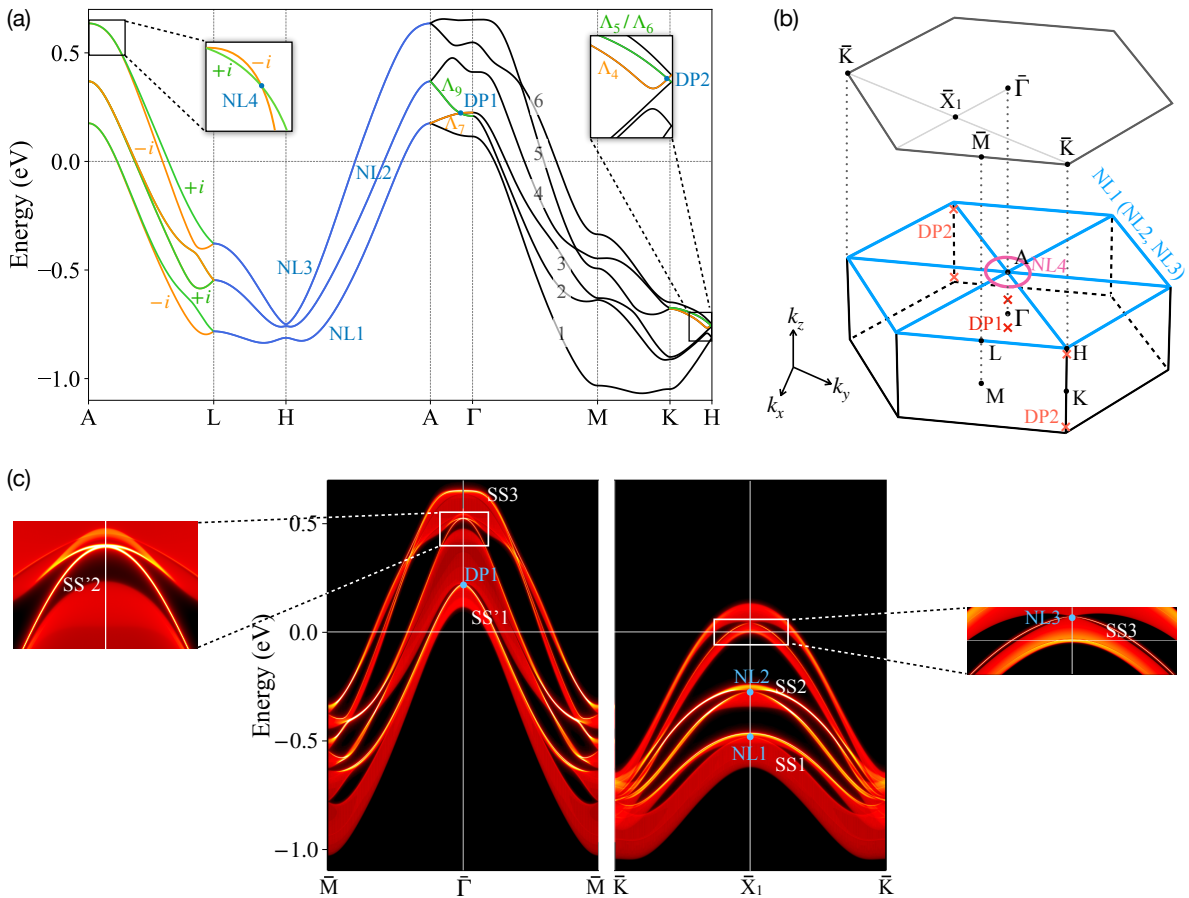}
    \caption{\label{Fig4} Electronic structure of Pb$_{1/3}$TaS$_2$. (a) Electronic bands along a high-symmetry path. Energy is measured from the Fermi level. Orange and green distinguish crossing bands with different symmetry operation eigenvalues or irreducible representations (IRs). Dirac line and point degeneracies are labeled and highlighted in blue. Note that along the $\Gamma$-$M$-$K$-$H$ line, all the seemingly gap closings are actually open at the meV scale except for DP2 and those at point $H$. Insets display zoom-in views of the band structure. The band gap in the left inset is exaggerated to enhance the visibility. (b) Brillouin zone (BZ) and surface BZ of Pb$_{1/3}$TaS$_2$. High-symmetry points are marked. Dirac nodal lines and points are indicated. (c) Momentum-resolved densities of states (DOS) of a semi-infinite (001) Pb-terminated slab. Yellow and dark red correspond to high and low density, respectively. The bulk nodes as well as surface states are highlighted.}
\end{figure*}

Fig. \ref{Fig2}(a) shows the in-plane resistivity ${\rho}_{ab}$ as a function of temperature $T$ for Pb$_{1/3}$TaS$_2$, which exhibits metallic behavior with a high residual resistivity ratio ($RRR = \rho(300 K)/\rho(3 K)$) amounting to 24.
The zoom-in figure in the inset of Fig. \ref{Fig2}(a) shows a sharp superconducting transition, $T_c^{50\%}$ of which determined at the half value of the normal state resistivity is around 2.8 K, higher than 2$H$-TaS$_2$ ($T_c$ = 0.8 K) \cite{NAGATA1992} and PbTaS$_2$ ($T_c$ = 2.6 K) \cite{PbTaS2}.
Fig. \ref{Fig2}(b) shows the $T$ dependence of dc magnetic susceptibility with an external magnetic field ($H$ = 2 Oe) along ab-plane.
The diamagnetic signal reveals a superconducting transition at 2.7 K in consistent with transport measurements. The shielding volume fraction of superconductivity is close to 100\% from the zero-field cooling (ZFC) process.

The Hall data $\rho_{yx}$ at different temperatures is presented in Fig. \ref{Fig2}(c). $\rho_{yx}$ is positive and linear in field, which suggests holes dominate in charge transport. The inset of Fig. \ref{Fig2}(c) presents the Hall coefficient $R_H=\rho_{yx}/H$, which varies slowly with $T$. Note that the carrier concentration $n$ cannot be directly deduced from $R_H$ given the multi-band nature of Pb$_{1/3}$TaS$_2$ (see the band structure below).

Fig. \ref{Fig2}(d)-(e) show the $T$ dependent ${\rho}_{ab}$ at various fields perpendicular and parallel to the ab-plane, respectively. The superconducting transition broadens and shifts towards lower temperatures by increasing fields, due to a field-induced pair breaking effect. The upper critical field for $H$//$ab$ ($H_{c2}^{ab}$) and $H$//$c$ ($H_{c2}^c $) versus $T_c^{50\%}$ are summarized in Fig. \ref{Fig2}(f), which exhibits an upward curvature. Similar features were reported in PbTaSe$_2$ \cite{PbTaSe2014} and PbTaS$_2$\cite{PbTaS2}, in which the upward feature was roughly fitted by
\begin{equation}\label{Eq1}
H_{(c2)}(t) = H_{(c2)}(0)(1-t^{3/2})^{3/2}
\end{equation}

where $t = T/T_c$. Eq. \ref{Eq1} arises from a local-pairing mechanism\cite{Hc21990}. However, Eq. \ref{Eq1} does not fit our data well.

An alternative interpretation of the upward feature suggests that Pb$_{1/3}$TaS$_2$ has multi-gap nature, as in MgB$_2$\cite{MgB22002Cp,MgB2Hc2}, NbSe$_2$\cite{NbSe2} and some iron-based superconductors\cite{Fe-based,SmFeAsO,Co-BaFe2As2}. As seen in Fig. \ref{Fig2}(f), the data is well fitted by a two-gap model\cite{twoband2013}:

\begin{equation}\label{Eq2}
\begin{aligned}
a_0[lnt + U(h)]&[lnt + U(\eta h)] + a_1[lnt + U(h)]
\\&+ a_2[lnt + U(\eta h)] = 0
\end{aligned}
\end{equation}

where $t=T/T_c$ and $h = H_{c2}D_1/(2\phi_0/T)$ is the reduced temperature and critical field, in which
 $a_0 = 2(\lambda_{11}\lambda_{22}-\lambda_{12}\lambda_{21})$, $a_1 = 1+(\lambda_{11}- \lambda_{22})/\lambda_0$, $a_2 = 1-(\lambda_{11}-\lambda_{22})/\lambda_0$, $\lambda_0 = ((\lambda_{11}-\lambda_{22})^2 +
4\lambda_{12}\lambda_{21})^{1/2}$,  $\eta= D_2/D_1$ and $U(x) = \psi(1/2 + x) -\psi(1/2)$. $\lambda_{11}$ ($\lambda_{22}$) and $\lambda_{12}$($\lambda_{21}$) are the intraband and interband BCS coupling constants, respectively , $\psi(x)$ is the digamma function, $D_1$ and $D_2$ are the diffusivity of each band.

According to the fitting, the upper critical field at zero-$T$ $H_{c2}^{ab}$(0) and $H_{c2}^c$(0) are estimated to be 6.84 T and 0.4 T, respectively.
The in-plane $H_{c2}^{ab} $(0) slightly exceeds the Pauli paramagnetic limit (${{\mu_0H}_P^{BCS}}$(0) = 1.84 $T_c$ $\approx 5.15$ T).
Subsequently, the superconducting coherence length is calculated to be ${\xi}_{ab}$(0) ${\approx}$ 28.7 nm and ${\xi}_c$(0) ${\approx}$ 1.68 nm through the Ginzburg-Landau(GL) formula: $H_{c2}^{c} = \Phi_0/2\pi\xi _{ab}^2 $ and $H_{c2}^{ab} = \Phi_0/2\pi\xi_{ab}\xi _{c} $. Interestingly, ${\xi}_{c}$(0) is close to the lattice constant. Moreover, the superconducting anisotropy ($\gamma_{H_{c2}}$ = $H_{c2}^{ab} /H_{c2}^{c}$) is as large as 17.1, larger than that in 2$H$-TaS$_2$ (6.7)\cite{TaS2Hc2}, which is consistent with the fact that the distance between adjacent TaS$_2$ blocks is 7.42 \r{A} for Pb$_{1/3}$TaS$_2$, larger than 6.05 \r{A} of 2$H$-TaS$_2$ \cite{MTaX21980}. Above all, the results indicate the q-2D superconducting nature in Pb$_{1/3}$TaS$_2$.

In order to gain further information of the superconducting state, the isothermal magnetization $M(H)$ with field along ab-plane is presented in Fig. \ref{Fig3}(a) at various temperatures. The inset shows the full magnetization loop at 1.8 K, indicating Pb$_{1/3}$TaS$_2$ is a typical type-II superconductor.

The in-plane lower critical field $H_{c1}^{ab}$ can be determined at the point which the magnetization curve starts to deviate from the linear Meissner response. $H_{c1}^{ab}$ versus $T$ is plotted in Fig. \ref{Fig3}(b), which is fitted by the conventional formula $H_{c1}(T)=H_{c1}(0)[1-(T/T_c)^2$] for a weakly coupled
superconductor\cite{Hc1}. $H_{c1}^{ab}$ at zero-$T$ is estimated to be 16.3 Oe.

Using the relationship $H_{c2}^{ab}(0)/H_{c1}^{ab}(0) = 2\kappa_{ab}^2/ln\kappa_{ab}$ and $\kappa_{ab}(0)= \lambda_{ab}(0)/{\xi}_{c}(0)$, we estimate the GL parameter $\kappa_{ab}$ $\sim$ 98.1 and penetration depth $\lambda_{ab}(0)$ $\sim$ 164.8 nm, the resulted parameters are summarized in Table \ref{Tab1} for brevity, indicating Pb$_{1/3}$TaS$_2$ is an extreme type-II superconductor with highly anisotropic properties.

\begin{table}[hpbt]
\centering
\caption{\label{Tab1} The anisotropic superconducting parameters of Pb$_{1/3}$TaS$_2$ single crystal.}
\setlength{\tabcolsep}{10mm}
\begin{tabular}{ccccc}
\hline
Parameters             & Values (unit)  \\
\hline
$H_{c2}^{ab}(0)$       & 6.84 (T)  \\
$H_{c2}^{c}(0)$  &   0.4 (T) \\
${\xi}_{ab}$(0)  & 28.7 (nm) \\
${\xi}_{c }$(0) &  1.68 (nm)\\
$H_{c1}^{ab}(0)$  &   16.3 (Oe) \\
$\lambda_{ab}$(0) &   164.8 (nm)\\
$\kappa_{ab}$  &   98.1\\
$\gamma_{H_{c2}}$ & 17.1 \\
\hline
\end{tabular}
\end{table}

\vspace{3ex}

\noindent\textbf{Band structures}
\noindent Now we report the topological electronic bands of the Pb$_{1/3}$TaS$_2$ system predicted by our first-principles calculations. Fig. \ref{Fig4}(a) presents the electronic band structure of   Pb$_{1/3}$TaS$_2$ in the presence of SOC, with high-symmetry \textit{k} points given in Fig. \ref{Fig4}(b). Six isolated bands (numbered as band 1 to 6 from low to high energy) are located in the vicinity of the Fermi energy. Each band has double degeneracy (Kramers pair) because of the $\mathcal{PT}$ symmetry. The band dispersion along the $k_z$ direction (i.e., $A$-$\Gamma$, $K$-$H$) is relatively weak in comparison to that along the in-plane direction (i.e., $A$-$L$-$H$-$A$, $\Gamma$-$M$-$K$), indicating the q-2D of the material, in agreement with the experimental observation.

For further analysis, it is essential to list the crystalline symmetries in Pb$_{1/3}$TaS$_2$ that are of particular importance: three vertical mirrors $\mathcal{M}_{[1\bar10]} = \left\lbrace\bar{2}[1\bar10]\right|\left.0,0,0\right\rbrace$, $\mathcal{M}_{[120]} = \left\lbrace\bar{2}[120]\right|\left.0,0,0\right\rbrace$, $\mathcal{M}_{[210]} = \left\lbrace\bar{2}[210]\right|\left.0,0,0\right\rbrace$, a horizontal mirror $\mathcal{M}_{[001]} = \left\lbrace\bar2[001]\right|\left.0,0,\frac12\right\rbrace$, a threefold rotation $\mathcal{C}_{3[001]} =  \left\lbrace3[001]\right|\left.0,0,0\right\rbrace$, and a sixfold screw $\widetilde{\mathcal{C}}_{6[001]} =  \left\lbrace6[001]\right|\left.0,0,\frac12\right\rbrace$. In combination with vertical mirrors and the $\mathcal{PT}$ symmetry, $\mathcal{M}_{[001]}$ nonsymmorphically dictates multiple Dirac nodal lines on the $k_z = \pi$ plane [colored by blue in Fig. \ref{Fig4}(b)] for every two bands \cite{PhysRevB.95.075135,PhysRevB.97.045131}. We refer to the nodal lines between bands 1 and 2 as NL1, similarly for NL2 and NL3, as illustrated in Fig. \ref{Fig4}(a). Crossing bands for each nodal line are distinguished by opposite $\mathcal{M}_{[001]}$ eigenvalues ($i$ or $-i$) as shown in Fig. \ref{Fig4}(a). Detailed argument can be found in Supporting Information. Apart from the symmetry-enforced degeneracies, an accidental $\mathcal{M}_{[001]}$-protected nodal line (NL4) exists between bands 5 and 6 [Fig. \ref{Fig4}(a) (left inset) and Fig. \ref{Fig4}(b)]. As stabilized by the mirror $\mathcal{M}_{[001]}$, each line node in NL1-4 is shown to carry a zero-dimensional topological charge $Q=1$ \cite{PhysRevB.95.075135}, confirming the nontrivial topology of above nodal lines.

Along with the line degeneracies, Fig. \ref{Fig4}(a)-(b) reveal two sets of Dirac points on the rotation/screw axes $A$-$\Gamma$ and $K$-$H$, denoted as DP1 and DP2, with the stability guaranteed by the sixfold screw $\widetilde{\mathcal{C}}_{6[001]}$ and threefold rotation $\mathcal{C}_{3[001]}$, respectively. If $2\%$ compression along the \textit{b} axis breaks $\mathcal{C}_{3[001]}$ and consequently opens up DP2, the crossing bands (bands 4 and 5) will become fully gapped, leading to a $\mathbb{Z}_2 = 1$. See Supporting Information for details. Therefore, topological surface states will emerge between bands 4 and 5 \cite{TI2010}, as discussed below.

By constructing a semi-infinite (001) slab with the Pb termination, we demonstrate the momentum-resolved surface density of states (DOS) of Pb$_{1/3}$TaS$_2$ in Fig. \ref{Fig4}(c). High-symmetry points in the reduced surface BZ are given in Fig. \ref{Fig4}(b). Surface states and the corresponding bulk line or point nodes are labeled. Along the $\bar{M}$-$\bar{\Gamma}$-$\bar{M}$ path [left panel of Fig. \ref{Fig4}(c)], surface states related to DP1, DP2 and NL3 are observed in turn from low to high energy. The surface band SS'2 is particularly illustrated in the left zoom-in view of Fig. \ref{Fig4}(c). SS'2 consistently appears even if DP2 is opened up by a symmetry breaking (e.g., compression along the \textit{b} axis), because of the inverted band structure and the resulting $\mathbb{Z}_2$ invariant 1. Along the $\bar{K}$-$\bar{X}_1$-$\bar{K}$ direction [right panel of Fig. \ref{Fig4}(c)], line nodes responsible for NL1, NL2 and NL3, as well as the induced boundary modes are emergent. We note that NL4 is covered by the high-intensity surface modes from NL3, therefore invisible in the surface spectrum.

\vspace{3ex}
\noindent\textbf{Discussions}

\noindent Although many candidates of NLSMs have been proposed, experimental realizations of NLSMs are still relatively scarce \cite{MSiS,fujioka2019strong}. One of the biggest challenges is that the most of the NLSMs theoretically predicted are fragile to SOC (e.g., PbTaS$_2$\cite{PbTaS2} and SnTaS$_2$ \cite{SnTaS2PRB}). In our study, we have proposed a new TCM Pb$_{1/3}$TaS$_2$ hosting nodal-line structures, which remain stable even under significant SOC by virtue of the nonsymmorphic symmetry protection, and hence can be observed in a realistic experiment.
Remarkably, Pb$_{1/3}$TaS$_2$ also possesses multiple Dirac points that can be driven into a TI phase by breaking the rotation symmetry. We predict Dirac-cone-like surface modes on the (001) surface, owing to the nontrivial band topology. Moreover, the q-2D feature and anisotropic transport are also supported by the band dispersions. Our results signify that Pb$_{1/3}$TaS$_2$ could act as a prospective platform to study the interaction between topological property and superconductivity. For the future work, angle-resolved photoemission spectroscopy (ARPES) and scanning tunneling microscopy (STM) studies are needed to identify the band structure and superconducting gap directly.

In summary, from first-principles calculations, the centrosymmetric Pb$_{1/3}$TaS$_2$ manifests multiple nodal states along with Dirac-cone-like surface states in the presence of SOC. Additionally, our experimental results suggest q-2D superconductivity with highly anisotropic features in this system. The combination of nontrivial band topology and superconductivity makes Pb$_{1/3}$TaS$_2$ a new candidate for further research of TSCs. Our work presents an important breakthrough in searching for new topological phases by building blocks design based on symmetry analysis.

\vspace{3ex}
\noindent\textbf{Methods}

\noindent\textbf{Sample preparation}
The Pb$_{1/3}$TaS$_2$ single crystals were prepared by the chemical vapor transport (CVT) method. Stoichiometric amounts of high-purity Pb, Ta, S powders with the transport agents PbBr$_2$ (10 mg/cm$^3$ in concentration) were thoroughly mixed and sealed in an evacuated quartz tube. The tube was heated at 1173 K with a temperature gradient of 5 K/cm for one week in a two-zone furnace.
\vspace{2ex}

\noindent\textbf{Measurements}
The X-ray diffraction (XRD) pattern was performed on a Bruker D8 Advance X-ray diffractometer with Cu-K$_{\alpha}$ radiation. The chemical composition was determined by an energy-dispersive x-ray (EDX) spectrometer (Model Octane Plus) affiliated to a Zeiss Gemini 450 Schottky field emission scanning electron microscope (SEM). The transport measurements were measured on an Oxford superconducting magnet system equipped with a $^3$He cryostat. The DC magnetization was carried out on a Quantum Design magnetic property measurement system (MPMS3).

\vspace{2ex}

\noindent\textbf{Band calculations}
The density functional theory (DFT) calculations are performed using the Vienna \textit{ab initio} simulation package ({\sc vasp}) \cite{PhysRevB.54.11169}, based on the generalized gradient approximation (GGA) method under the Perdew-Burke-Ernzerhoff (PBE) parameterization \cite{PhysRevLett.77.3865}. The energy cutoff of the plane wave is 323.4 eV. The Brillouin zone (BZ) is sampled by a $12\times12\times4$ grid for the self-consistent calculations. Irreducible representations (IRs) of electronic eigenstates at high-symmetry \textit{k}-points are determined via an in-house code and the software package \texttt{irvsp} \cite{gao2021irvsp}. Wannier functions are constructed by projecting Bloch states onto Ta $5d$ orbitals through {\sc wannier90} \cite{PhysRevB.56.12847, PhysRevB.65.035109, pizzi2020wannier90} without the iterative maximal localization procedure. Nodal lines or points and surface spectrum are computed with the {\sc wanniertools} package \cite{wu2018wanniertools}, where the latter is based on the iterative Green's function method \cite{sancho1985highly}. Pre/Post-processing tools and utilities for solids computation \cite{wang2019vaspkit, momma2008vesta, setyawan2010299, togo2018texttt} are exploited.
%
\vspace{3ex}

\noindent\textbf{Data availability}

\noindent The data that support the findings of this study are available from the corresponding author upon reasonable request.

\vspace{3ex}

\noindent\textbf{Acknowledgments}

\noindent The authors are grateful to Chao Cao, Wei Zhu and Takuya Nomoto for helpful discussion. C.X. acknowledges the computational resources at the HPC center at Hangzhou Normal University in China and the RIKEN Center in Japan.
This research was supported by the National Natural Science Foundation of China via Project 11904294 and 11774305, National Key Projects for Research \& Development of China (Grant No. 2019YFA0308602), Zhejiang Provincial Natural Science Foundation of China under Grant No. LQ19A040005 and the foundation of Westlake Multidisciplinary Research Initiative Center (MRIC)(Grant No. MRIC20200402). M.H. acknowledges the support from JST CREST (Grants No. JPMJCR19T2).
The authors thank the support provided by Dr. Xiaohe Miao and Dr. Lin Liu from Instrumentation and Service Center for Physical Sciences at Westlake University.

\vspace{3ex}

\noindent\textbf{Author contributions}

\noindent X.Y. conceived the project and grew the Pb$_{1/3}$TaS$_2$ single crystals, X.Y. performed the characterization and analyzed the data with the help of J.W., W.H., Z.X., T.W. and C.Z.. T.Y. and C.X. performed the first-principles calculations and analyzed the electronic structure. X.Y., T.Y. and X.L. wrote the manuscript with contributions from all authors.

\vspace{3ex}

\noindent\textbf{Additional information}

\noindent\textbf{Competing interests:} The authors declare no competing interests.
%

\end{document}